\documentclass[a4paper,aps,pra,showpacs,preprintnumbers,twocolumn]{revtex4}
\usepackage{graphics,graphicx,epsfig}
\usepackage[centertags]{amsmath}
\usepackage{amsfonts}
\usepackage{amssymb}
\usepackage[cp1251]{inputenc}
\usepackage[english]{babel}
\inputencoding{cp1251}
\usepackage{graphicx}
\usepackage{amsthm,color}
\usepackage{epstopdf}
\DeclareGraphicsExtensions{.pdf,.png,.jpg,.eps}

\begin{document}

\def\BE{\begin{equation}}
\def\EE{\end{equation}}
\def\BEA{\begin{eqnarray}}
\def\EEA{\end{eqnarray}}
\def\BY{\begin{eqnarray}}
\def\EY{\end{eqnarray}}

\def\L{\label}
\def\nn{\nonumber}
\def\ds{\displaystyle}
\def\o{\overline}

\def\({\left (}
\def\){\right )}
\def\[{\left [}
\def\]{\right]}
\def\<{\langle}
\def\>{\rangle}

\def\h{\hat}
\def\hs{\hat{\sigma}}
\def\td{\tilde}

\def\k{\mathbf{k}}
\def\q{\mathbf{q}}
\def\r{\vec{r}}
\def\ro{\vec{\rho}}
\def\a{\hat{a}}
\def\b{\hat{b}}
\def\c{\hat{c}}
\def\h{\hat}

\title{Effect of saturated absorption on sub-Poissonian lasing}


\author{Yu. M. Golubev$^{1}$, T. Yu. Golubeva$^{1}$, E. A. Vashukevich$^{1}$, S. V. Fedorov$^{2,3}$, N. N. Rosanov$^{2,3,4}$ }
\address{$^{1}$Saint-Petersburg State University, St. Petersburg, 199034, Russia,\\ $^{2}$Vavilov State Optical Institute, St. Petersburg, Russia,\\
$^{3}$ITMO University, St. Petersburg, Russia\\ $^{4}$Ioffe Institute, St. Petersburg, Russia }

\begin{abstract}
This paper discusses the effect of saturated absorption on the quantum features of a sub-Poissonian laser. It is shown that when both media (laser
and absorbing) are saturated, the appearance of an additional element in the cavity does not lead to the demolition of the quantum properties
(regular statistics of photons and quadrature squeezing) of laser radiation.
\end{abstract}


 \maketitle


\section{Introduction}

Lasers with saturable absorption first considered more then 50 years ago as generators of giant radiation pulses (Q-switching mode) \cite{1*,2*,3*},
have now found widespread use for other purposes, including ensuring passive mode synchronization of the laser \cite{4*}. Non-linear absorption
spectroscopy and radiation frequency stabilization using such lasers are another application areas \cite{Letohov1967, Chebotaev1968}. The detection
of a bistable lasing regime in such systems \cite{Kazantsev1970, Scott1975, Salomaa1973, Lee1968, Lugiato1978} ensured a renewed focus on lasers with
absorbing cell. Finally, the detection of optical spatial dissipative solitons in wide-aperture (multimode) laser with a saturable absorption
\cite{RF*,Rosanov2011} attracts the attention of researchers to such systems today.

Quantum noise in a laser with a saturable absorption began to be studied 40 years ago \cite{Lugiato1978, Roy1979}. However, the systems under
discussion are distinguished by a large variety of different laser generation regimes, so that the quantum noise study held today only for some of
them. In particular, the analysis of the solution of the Focker-Planck equation allowed the authors \cite{Lugiato1978} to predict the increase of
intensity fluctuations in the regions of bistability near the generation thresholds. In \cite{Roy1979}, on the basis of the density matrix
representation for a single-mode laser with a saturable absorption, probability distributions of the photon number were obtained. However, a large
complex of spatial aspects of quantum correlations, the study of the effects of squeezing of quantum field noise, remained undisclosed.

The presence of dissipative spatial solitons in the dynamics of laser radiation with the absorbing cell unwittingly forces us to draw parallels
between these systems and others, where similar localized structures of the electromagnetic field were also observed. In this regard, the most widely
investigated systems are the optical parametric oscillator and the interferometer with the Kerr medium. The study of the quantum properties of
spatial solitons in these systems revealed a wide variety of quantum effects. Thus, the work \cite{Kolobov-book} discusses the full amplitude
squeezing (squeezing in the whole beam) for generated spatial solitons and the authors \cite{Kolobov-book, Fabre2000} indicate the possibility of
observing local quantum correlations in spatial solitons and point to the presence of quantum anti-correlations between different transverse parts of
the soliton, which are formed as a result of diffraction. The influence of quantum fluctuations on the position of the center of a spatial soliton
was investigated in \cite{Nagasako1998}. The authors of \cite{Mecozzi1998} note that the use of simple diaphragm (cutting out the central part of the
soliton beam) provides intensity squeezing of the remaining light.

For a wide-aperture interferometer with a Kerr nonlinearity irradiated by external coherent light (plane wave), \cite{BR*} showed the possibility of
suppressing of quantum fluctuations. A consistent theory of quantum fluctuations of spatial dissipative optical solitons in such an interferometer is
developed in \cite{NKKR*, NVR_1*, NVR_2*, Rosanov2011}.

The discovery of all these non-trivial quantum features of dissipative spatial solitons generated by related systems forces us to turn again to the
analysis of the properties of laser radiation with saturable absorption already from these positions. This paper is the first of a series of papers
devoted to this study.

Our interest was attracted by the question of the conservation of quantum field statistics in the presence of an absorbing cell. It is well known
that quantum statistics of a radiation are sensitive to any uncontrolled losses. We consider here the situation when inside the cavity there are two
media: the active medium of the laser with a partially regular pump and the absorbing cell. It is well known that regular pump in the absence of an
absorbing medium forms the sub-Poissonian statistics of lasing \cite{GolSok}. It can be expected that the processes of absorption, being random in
time, will destroy the regularity, and hence destroy the squeezing of the field according to the number of photons. However, we will demonstrate that
such intuitive reasoning turns out to be incorrect, and we will discuss the physical reasons for the conservation of quantum features of a field.

The paper is organized as follows. Section II presents the derivation of the equation for the lasing field in the adiabatic approximation in the
presence of an additional absorbing cell. The condition of the stability of lasing is discussed. In Section III, the problem of photon statistics and
phase diffusion of lasing with saturated absorption is considered. Regular pumping of the active medium and the lack of synchronization are assumed.
Section IV considers lasing with saturated absorption with synchronization of generation by an external field. The effect of the quadrature field
squeezing is discussed.

\section{Equation for single-mode lasing with saturated absorption}

\subsection{Initial Heisenberg-Langevin equations for active and passive media}

In this paper, a sub-Poissonian laser with an absorbing cell will be discussed. Strictly speaking, this system has already been considered under the
quite good quantum-electrodynamic approach \cite{Lugiato1978}. However, the main focus was not on the quantum properties of the generation field, but
on the effects of bistability.

We are mainly interested in the question of how much the quantum features of a sub-Poissonian laser are able to conserve under conditions when an
additional element in the form of a resonantly absorbing cell is placed inside the cavity.

Formally, we will follow the work \cite{GIG}, where the theory of a sub-Poissonian laser, examined under the quantum electrodynamics, under
conditions of phase-locking by a weak external field in a coherent state is considered. The equations obtained in \cite{GIG} and generalized to the
case of the presence of additional resonance absorption in the system will be taken as a basis.

To describe both multiatomic media (active, on the basis of which lasing is carried out, and passive, realizing saturated absorption), collective
variables are used. For the active medium, these are the coherence operators $\hat \sigma$, the upper-level population  $ \hat \sigma_2 $ and the
lower-level population $ \hat \sigma_1 $. A similar set of operators also holds for the passive medium: $\hat \pi $ (coherence), $\hat\pi_2$
(upper-level population) and $\hat \pi_1$ (lower-level population).

All Heisenberg equations for collective variables of media can be written in a standard way based on the Hamiltonian
\BY
 &&\hat H=\[\hbar\omega_0  (\hat a^\dag\hat a+1/2)\]\nn\\
&& +\[W_2\;\hat\sigma_2+ W_1\;\hat\sigma_1\]+
 \[W_{p2}\;\hat\pi_2+ W_{p1}\;\hat\pi_1\]\nn\\
&&+\[i\hbar g (\hat\sigma\hat a^\dag-\hat\sigma^\dag\hat
a)\]+\[i\hbar g_p (\hat\pi\hat a^\dag-\hat\pi^\dag\hat a)\].\L{1}
\EY
Here, the first three terms on the right describe an independent laser field with an amplitude of $\hat a$ and active and passive media. The last two
terms describe the interaction of the laser field with both media.
\begin{figure}
\includegraphics[scale=0.5]{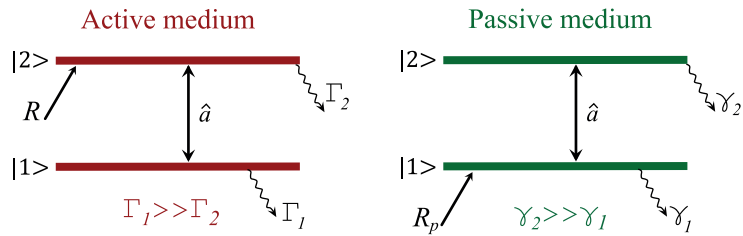}
\caption{ The schematic draw of the active and passive mediums.} \L{fig1}
\end{figure}

It should be noted that here we choose all frequency detunings equal to zero (see fig.\ref{fig1}). In all obtained equations for the field and media
relaxation terms and the corresponding Langevin noise sources are introduced. Such equations are called Heisenberg-Langevin equations.

It seems to us the most convenient is to go to the c-number analogue of the Heisenberg operator amplitudes. It is important to emphasize that this
transition is well formalized and, although the operator amplitudes become non-operator when implemented, this does not mean that we return to
classical electrodynamics. At the same time, the difficulties associated with this approach are well known. The equations give us the opportunity to
calculate only normally ordered means.

The Heisenberg-Langevin equations in the c-number representation for the active medium coincide with those obtained in \cite{GIG}, and have the
following form: \BY
&&\dot{\sigma}=-\Gamma_1/2\;\sigma+g\[\sigma_2-\sigma_1\]
a+ F,\L{2}\\
&&\dot{\sigma}_1=-\Gamma_1 \sigma_1+g\[\sigma^\ast
a+\sigma a^\ast\]+ F_1,\\
&&\dot{\sigma}_2=R-\Gamma_2\sigma_2-g\[\sigma^\ast a+
a^\ast\sigma\]+ F_2.\L{4}
\EY
Here $R$ is the average rate of excitation of the active medium. We also assumed that $\Gamma_2\ll\Gamma_1$, which provides the best conditions for
sub-Poissonian lasing \cite{GolSok}. In this case, the coherence decay rate is $(\Gamma_1 + \Gamma_2)/2\approx\Gamma_1/2 $.

Langevin sources in the equations (\ref{2}) - (\ref{4}) are given by non-zero cross-correlators:

\BY
&&\langle F^\ast(t)
F(t\;^\prime)\rangle=\[\Gamma_1\o\sigma_2+R\]\delta(t-t^\prime),\L{5}\\
&&\langle F(t) F(t\;^\prime)\rangle=2g\;\o{\sigma a}\;\delta(t-t^\prime),\\
&&\langle F_1(t)
F(t\;^\prime))\rangle=\Gamma_1\o\sigma\delta(t-t^\prime),\\
&&\langle F_2(t) F_2(t\;^\prime)\rangle\nn\\
&&=\[\Gamma_2\o\sigma_2+R(1-s)-g\(\o{ a^\ast\sigma}+\o{a\sigma^\ast}\)\]\delta(t-t^\prime),\\
&&\langle F_1(t) F_1(t\;^\prime)\rangle=\[\Gamma_1\o\sigma_1-
g\(\o{ a^\ast\sigma}+\o{a\sigma^\ast}\)\]\delta(t-t^\prime),\quad\\
&&\langle F_2(t) F_1(t\;^\prime)\rangle=
g\(\o{a^\ast\sigma}+\o{a\sigma^\ast}\)\delta(t-t^\prime).\L{10}
\EY
Further we will apply the adiabatic approximation, through which we will be able to eliminate all atomic variables from the formal scheme and then
analyze the only closed equation for the field amplitude. As is known, this is possible under the condition that relaxation processes in media occur
much faster than the lasing field develops. This means that we must require the fulfillment of inequalities: $\kappa\ll\Gamma_1,\;\Gamma_2 $.

For further calculations, we will need solutions of the averaged equations (\ref{2}) - (\ref{4}) under stationary conditions. Assuming after
averaging $\dot\sigma=\;\dot\sigma_1=\;\dot\sigma_2=0$, we get \BY
&&\o\sigma_1=\frac{R}{\Gamma_1}\frac{I}{1+I},\quad\o\sigma_2=\frac{R}{\Gamma_2}\frac{1}{1+I}, \quad g\o\sigma=\frac{1}{2}\; \frac{\beta
Ra}{1+I},\quad \nn\\
&& I=\beta|a|^2,\quad\beta=\frac{4g^2}{\Gamma_1\Gamma_2} .\L{11}
\EY
One can note that the correlation function $ \langle F_2(t)F_2(t^\prime)\rangle$ turns out to depend on the statistical parameter $0<s<1$, which
determines the excitation mechanism of the laser medium. With a completely random (Poissonian) pump, $s = 0$, and with a strictly regular
(sub-Poissonian) pump, $s = 1$.

With regard to the passive medium, it is also two-level and similar in structure to the active one. But its set of parameters
$\gamma_1,\;\gamma_2,\;g_p,\;R_p$, generally speaking, does not coincide with the corresponding set for the active medium
$\Gamma_1,\;\Gamma_2,\;g,\;R$.

We choose level structures such that the equations for the passive medium are obtained from the equations for the active medium (\ref{2})-(\ref{4})
by simultaneously replacing $\Gamma_1 \leftrightarrows\Gamma_2 $ and $ \sigma_1 \leftrightarrows \sigma_2 $. In particular, this means that $
\gamma_1 \ll \gamma_2 $. Using this symmetry, we obtain the equations
\BY
&&\dot{\pi}=-\gamma_2/2\;\pi+g_p\[\pi_1-\pi_2\]
a+ G,\L{12}\\
&&\dot{\pi}_1=R_p-\gamma_1\pi_1-g_p\[\pi^\ast a+ a^\ast\pi\]+
G_1,\\
&&\dot{\pi}_2=-\gamma_2 \pi_2+g_p\[\pi^\ast a+\pi a^\ast\]+
G_2.\L{14}
\EY

Similarly, from (\ref {5})-(\ref{10}) we obtain the correlation functions for Langevin sources of the passive medium
\BY
&&\langle G^\ast(t)
G(t^\prime)\rangle=\[\gamma_2\o\pi_1+R_p\]\delta(t-t^\prime),\L{15}\\
&&\langle G(t) G(t^\prime)\rangle=2g_p\;\o{\pi}\;\delta(t-t^\prime),\\
&&\langle G_2(t)
G(t^\prime))\rangle=\gamma_2\o\pi\delta(t-t^\prime),\\
&&\langle G_1(t) G_1(t^\prime)\rangle \nn\\
&&=\[\gamma_1\o\pi_1+R_p(1-s_p)-
g_p\(\o{ a^\ast\pi}+\o{a\pi^\ast}\)\]\delta(t-t^\prime),\\
&&\langle G_2(t) G_2(t^\prime)\rangle=\[\gamma_2\o\pi_2-
g_p\(\o{ a^\ast\pi}+\o{a\pi^\ast}\)\]\delta(t-t^\prime),\quad\quad\\
&&\langle G_1(t) G_2(t^\prime)\rangle=
g_p\(\o{a^\ast\pi}+\o{a\pi^\ast}\)\delta(t-t^\prime).\L{20}
\EY
We assume that, unlike the active medium, the passive medium is excited to the lower level only with Poissonian statistics, that is, $s_p=0$ is
always the same (in contrast with  the similar parameter $s$, which may not be equal to zero).

As well as for the active medium, in the equations (\ref{12}) - (\ref{14}) it is necessary to put $\dot\pi, \; \dot\pi_1,\;\dot\pi_2=0$ and perform
averaging. Then \BY
&&\o\pi_2=\frac{R_p}{\gamma_2}\frac{I_p}{1+I_p},\quad\o\pi_1=\frac{R_p}{\gamma_1}\frac{1}{1+I_p}, \quad g_p\o\pi=\frac{1}{2}\;
\frac{R_p\beta_p}{1+I_p},\nn\\
&& I_p=\beta_p|a|^2,\quad\beta_p=\frac{4g_p^2}{\gamma_1\gamma_2}.\L{21}
\EY

\subsection{Heisenberg-Langevin equation for lasing field}

As one can see, all the equations of the medium (\ref{2}) - (\ref{4}) and (\ref{12}) - (\ref{14}) depend on the amplitude of the lasing field. In
order to close the system of equations, we must add to our consideration the equation for the c-number amplitude of the field $a (t)$. Taking into
account the fact that two media (active and passive) interact with the generation field, it can be written as follows: \BY
&&\dot{  a}=-{\kappa}/{2}\;(a-a_{in}) +g\sigma+g_p \pi.\L{22}
\EY
Here, the second and third terms on the right describe the effect on the field of generation of the active and passive components of the atomic
medium. This effect is realized through the collective coherence of the active medium $\sigma$ and through the collective coherence of the passive
medium $\pi$ \cite{DAVID}.

The first term on the right describes two processes. First, the escape of the laser field at the speed $\kappa$  from the cavity is related to the
transparency of the output mirror of the cavity. Second, the incoming of a weak signal in a coherent state with an amplitude of $a_ {in}$ from the
outside.

The presence of this signal is crucial for the occurrence and observation of such a quantum phenomenon as quadrature squeezing. The signal should be
limited in magnitude above and below. On the one hand, it should be weak enough so that the quantum features, if they arise in lasing, would not be
suppressed by it. On the other hand, the suppression of phase diffusion (phase locking) is provided by a signal with sufficient intensity. In
\cite{GIG}, this issue was investigated, and it was shown that these two restrictions are consistent.

At the same time, there is no need to apply synchronization if we want to follow the statistics of photons in lasing. As is known, in this case,
phase diffusion takes place, but it does nothing to prevent the occurrence of the sub-Poissonian statistics of photons.

Next, we will consider two variants of lasing, namely, with synchronization by an external field with amplitude $ a_{in} $ and without
synchronization, that is, with $ a_{in} = 0 $.

As known \cite{GolSok}, in the absence of an absorbing cell and with regular excitation of the active medium (for $s=1$), the photon statistics turns
out to be sub-Poissonian. To observe this quantum feature, there is no need to lock the lasing phase, since this property does not depend on the
phase at all. With regard to the additional absorption cell, we must understand whether it is possible to minimize its negative influence on the
quantum features of the field.

A more complicated situation arises if one wants to observe another quantum effect, namely, the quadrature squeezing of the field. To make this
possible, you need to "stop" \; phase diffusion in the laser field, which is ensured by introducing into consideration the external field with
amplitude $a_ {in}$ and a frequency that coincides with the carrier frequency of the laser field without phase locking \cite{GIG}.

\subsection{Adiabatic elimination of atomic variables}

In order to be able to eliminate adiabatically the atomic media, we should assume that all the atomic variables $ (\sigma, \; \sigma_1, \; \sigma_2)
$ and $ (\pi, \; \pi_1,\; \pi_2 ) $ evolve much faster than the field amplitude $a$. This assumption takes place under condition that $ \kappa $ is
the smallest among all the peculiar parameters of the system, that is, $\kappa \ll\Gamma_1,\; \Gamma_2, \; \gamma_1, \; \gamma_2$. In this case, both
atomic media quickly reaches the equilibrium, even though the change of the field is not significant. This enables us to focus exclusively on
stationary solutions of the equations (\ref{2}) - (\ref{4}) and (\ref{12}) - (\ref{14}), which can be ultimately substituted into the equation for
field amplitudes (\ref{22}). As a result, we reformulate the theory of lasing with an absorbing cell in the form of a single closed differential
equation for the c-number field amplitude:
\BY
\dot a(t)
 =&&-\kappa/2\;(a-a_{in}) \nn\\
&& +\frac{1}{2}\(\frac{A}{1+\beta|a|^2}-\frac{A_p}{1+\beta_p|a|^2}\)a+\Phi+\Phi_p.\qquad\L{23}
\EY
Here $ A =\beta R$ is the linear gain of the active medium, and $A_p= \beta_p R_p$ is the linear absorption coefficient of the passive medium.

The Langevin sources $\Phi$ and $\Phi_p$ due to adiabatic procedures are explicitly expressed in terms of the original sources $ (F, \; F_1, \; F_2)
$ and $ (G, \; G_1, \; G_2) $. The latter can be found respectively in the equations (\ref {2}) - (\ref{4}) and (\ref{12})-(\ref{14}). Non-zero
correlation functions for new noise sources are given by
\BY
&&a^\ast a^\ast\langle\Phi\Phi\rangle= -\frac{|a|^2}{2}\frac{I}{1+I}\frac{A}{1+I}(1+s/2),\qquad\L{24}\\
&&\langle\Phi\Phi^\ast \rangle= -\frac{1}{2}\frac{I}{1+I}\frac{A}{1+I}(1+s/2)+\frac{1}{2}
\frac{A}{1+I},\\
&&a^\ast a^\ast\langle\Phi_p\Phi_p\rangle=-\frac{|a|^2}{2}\frac{I_p}{1+I_p}\frac{A_p}{1+I_p} ,\qquad\\
&&\langle\Phi_p\Phi_p^\ast \rangle=-\frac{1}{2}\frac{I_p}{1+I_p}\frac{A_p}{1+I_p}+\frac{1}{2} \frac{A_p}{1+I_p}.\L{25}
\EY

\subsection{The equation for the laser field in the classical limit}

The equation in the classical limit is obtained from the Eq. (\ref{23}) by removing Langevin noise sources, therefore it takes the form
\BY
&&\dot a(t)
 =-\kappa/2\;(a-a_{in}) +\frac{1}{2}\(\frac{A}{1+\beta|a|^2}-\frac{A_p}{1+\beta_p|a|^2}\)a.\nn\\
&& \L{26}
\EY
For stationary conditions, that is, for $ \dot a=0$, we obtain the equality
\BY
&&\kappa(\tilde a-a_{in })=  \(\frac{A}{1+\tilde I}-\frac{A_p}{1+\tilde I_p}\) \tilde a,\qquad\L{27}\\
&& \mbox{where} \qquad \tilde I=\beta|\tilde a|^2,\qquad \tilde I_p=\beta_p|\tilde a|^2.\nn
\EY
Here we denote the stationary semiclassical solution as $\tilde a$ or $\tilde I,\;\tilde I_p$.

Assuming that $a_ {in}=\sqrt{n_{in}}e^{i\varphi_{in}}$, it is not difficult to see that the stationary classical solution that satisfies equality
(\ref{27}) is
\BY
&&\tilde a=\sqrt{\tilde n}e^{i\varphi_{in}}.\L{28}
\EY

The influence of an external field with an amplitude $a_ {in}$ on this solution is twofold. First, it unambiguously fixes (“locks”) the phase of the
lasing field, and in the absence of an external field, this phase is completely arbitrary. Secondly, strictly speaking, the quantity $\tilde n$ turns
out to depend on $n_ {in}$. However, we can neglect this dependence by assuming $\mu=\sqrt{n_ {in}/\tilde n}\ll1$.
According to this assumption, we can conclude that in the analysis of the stability of a stationary solution, the factor $\mu $ can be ignored,
setting it equal to zero. Due to the top constraint, the external synchronization field cannot impose its (Poissonian) statistics on the intracavity
field.
Note that in order for the external field to lock the phase of the lasing field, it is also necessary to impose on its value the bottom constraint:
it suffices to require that the value of $n_{in}$ exceeded $1/\tilde n$ \cite{GIG}.

In the following sections, we will linearize the equation for the lasing field, assuming that the fluctuations of the field are small in comparison
with the classical solution. To make this possible, we should make sure that the stationary solutions of the equation (\ref{26}) are stable.

A detailed calculation of stability will not be given here. It is carried out in a standard and well-known way based on the Eq. (\ref{26}) through
the study of small deviations from the stationary solution.

Due to this procedure, it can be obtained that the requirement of stability of a stationary solution imposes the following restriction on parameter
of cooperativity of the passive media $A_p/\kappa$:
\BY
&&A_p/\kappa>(1+I_p)^2\frac{\beta}{\beta_p-\beta}=(1+\tilde
I_p)^2\frac{\tilde I}{\tilde I_p-\tilde I}.\L{29}
\EY
One can see, that with $\beta<\beta_p $ the inequality holds only for sufficiently large values of the parameter of cooperativity. However, when
$\beta_p<\beta $, the value on the right becomes negative and this requirement is waived.

\section{Sub-Poisson lasing with saturated absorption (without synchronization)}
\subsection{Zero approximation for lasing fluctuations}

Now let us return to the Eq.(\ref{23}), which differs from the classical one (\ref{26}) by the presence of Langevin sources $\Phi $ and $\Phi_p$. In
this section, we assume that the lasing is not synchronized with an external field, that is, $n_{in}=0 $. It is more convenient instead of  Eq.
(\ref{23}) for complex amplitude, to discuss two separate equations for amplitude and phase. We represent the complex c-number field amplitude as
$a=\sqrt n\exp(i\varphi) $ and rewrite the equation (\ref{23}) for variables $n$ and $\varphi$:
\BY
&&\dot n=-\kappa n +\(\frac{A}{1+\beta n}-\frac{A_p}{1+\beta_p n}\)n+\Phi_n,\L{30a}\\
&&\dot\varphi=\Phi_\varphi\L{30b}
\EY
Here, instead of the Langevin sources $\Phi $ and $\Phi_p $, which appeared in the Eq. (\ref{23}), new sources have arisen. These new sources related
to initial as follows:
\BY
&&\Phi_n=\[a^\ast \Phi+a \Phi^\ast\]+\[a^\ast \Phi_p+a \Phi_p^\ast\],\L{31a}\\
&&\Phi_\varphi=-\frac{i}{n}\[a^\ast \Phi-a^\ast \Phi\]-\frac{i}{n}\[a^\ast \Phi_p-a^\ast \Phi_p\].\L{31b}
\EY

Given the formulas (\ref{24}) - (\ref{25}), we can necessarily find all desired correlation properties for these sources.

The question of the statistics of photons in lasing is solved on the basis of the equation (\ref{30a}). It is essentially non-linear, and for its
solution, it is necessary to find an acceptable linearization procedure. It is usually assumed that fluctuations in the number of photons are small
in comparison to the stationary semiclassical solution, that is, the solution to the equation (\ref {30a}) is written as
\BY
&&n(t)=\tilde n+\varepsilon(t),\qquad\tilde n\gg|\varepsilon(t)|.\L{32}
\EY
For $ \mu = 0 $, the condition of stationary lasing with an absorbing cell is
\BY
&&\kappa =\frac{A}{1+\beta\tilde n}-\frac{A_p}{1+\beta_p\tilde
n}\quad (\tilde I=\beta\tilde n,\;
\tilde I_p=\beta_p\tilde n).\L{33}
\EY

The most interesting regime for the purposes of quantum optics is the regime of saturation of both media, which is realized under the condition $
(\tilde I, \; \tilde I_p\gg1)$. In this case, it seems that the role of the absorbing cell is negligible, and the second term on the right can be set
to zero. It is not difficult to verify that this is not the case. Indeed, if the impact of the absorbing cell is neglected, then from the equation
(\ref{33}) one can find that the number of photons $n$ is equal to $ R/\kappa$. However, this is valid only when $R_p\ll R $. One can verify this by
writing the exact formula:

\BY
&&\tilde n=R/\kappa-R_p/\kappa
.\L{}
\EY
\subsection{Photon statistics and phase diffusion in lasing}
Carrying out the procedure of linearization of the Eq. (\ref{30a}) by fluctuations of $ \varepsilon $, we obtain a simple equation of the form
\BY
 \dot\varepsilon
 =-D\;\varepsilon +\Phi_n.\L{34}
\EY
Here, the value of $D$ with its positive values acts as the decay rate of photon fluctuations, and, taking into account the relation (\ref{33}), is
written explicitly as follows:
\BY
D=\kappa\;\frac{\tilde I}{1+\tilde I}+\frac{A_p}{1+\tilde
I_p}\(\frac{\tilde I}{1+\tilde I}-\frac{\tilde I_p}{1+\tilde
I_p}\) .\L{35}
\EY

Applying the Fourier transform to the Eq. (\ref{34}), we can obtain the spectral noise power of lasing, expressed in terms of the spectral noise
power of a Langevin source:
\BY
&&\langle|\varepsilon_\omega|^2\rangle=\frac{\langle|\Phi_{n\omega}|^2\rangle}{D^2+\omega^2}.\L{36}
\EY
Using the formulas (\ref{31a}), (\ref{24}) - (\ref{25}) and (\ref{33}), we can find an explicit formula for the spectral power of the source noise:
\BY
&& \langle |\Phi_{n\omega}|^2\rangle=2\tilde nD_1,\L{37}\\
 && D_1=\kappa\;\frac{1-s/2\;\tilde I}{1+\tilde I}+
 \frac{A_p}{1+\tilde I_p}\(\frac{1-{s}/{2}\;\tilde I}{1+\tilde I}+\frac{1}{1+\tilde I_p}\).\nn
\EY
The spectral noise power of a Langevin source, and hence the spectral lasing power, depends on the method of excitation of the active medium, which
is determined by the parameter $s$. The most significant case is when both media are saturated, that is $\tilde I,\;\tilde I_p\gg1$. In this case,
the coefficients $ D $ and $ D_1 $ get a simple form:
\BY
 &&D=\kappa,\qquad D_1=-s/2\;\kappa.\L{38}
\EY
As one can see, in this case the values $\langle|\Phi_{n\omega}|^2\rangle $ and $ \langle|\varepsilon_\omega|^2\rangle $ turn out to be negative,
which indicates the presence of nonclassical features in the system.

Let us follow the signal, that will be observed in direct photodetection of light as it escapes from the cavity. As is well known \cite{GolSok}, the
spectrum of the photocurrent can be given by
\BY
&&\langle|\delta i_\omega|^2\rangle=\kappa|\tilde a|^2 +\kappa^2
\langle|\varepsilon_\omega|^2\rangle.\L{39}
\EY
Substituting the formulas (\ref{36}) and (\ref{37}), we get
\BY
&&\langle|\delta i_\omega|^2\rangle/\langle i\rangle=1+ \frac{2\kappa D_1}{D^2+\omega^2},\qquad \langle i\rangle=\kappa|\tilde a|^2. \L{40}
\EY
In regime of the both media saturation
\BY
&&\langle|\delta i_\omega|^2\rangle/\langle
i\rangle=\frac{\kappa^2(1-s)+\omega^2}{\kappa^2+\omega^2}.\L{41}
\EY
As can be seen, at $s = 1$ in the saturation regime, around zero frequencies the shot noise of a photocurrent is effectively suppressed. This result,
well known for lasing without an absorbing cell, is retained in the presence of saturated absorption.

Let us follow the changes of the phase of the lasing field. We recall that according to the semiclassical theory, the phase has a random value and
thus remains indefinite. To understand how quantum theory relates to this, we need to return to the Eq. (\ref{30b}). It is easy to get from
Eq.(\ref{30b}) the ratio between the two phases of the field at times $t_1$ and $t_2$:
\BY
&&\langle\[\varphi(t_2) -\varphi(t_1)\]^2\rangle\L{43}\\
&&=\iint_{t_1}^{t_2} dt dt^\prime \langle\Phi_\varphi(t) \Phi_\varphi(t^\prime)\rangle=\langle\Phi_\varphi^2\rangle\;(t_2-t_1).\nn
 \EY
Here we have taken into account that
\BY
&&\langle\Phi_\varphi(t)\Phi_\varphi(t^\prime)\rangle=\langle\Phi_\varphi^2\rangle\;\delta(t-t^\prime),\L{44}\\
&&\langle\Phi_\varphi^2\rangle=\frac{\kappa}{\tilde n}\(1+\frac{2A_p/\kappa}{1+\tilde I_p}\).\nn
\EY
The last formula is obtained from the definition (\ref{31b}) with regard to the equalities(\ref{24}) - (\ref{25}). When discussing the stationary
generation condition (\ref{33}), we came to the conclusion that in the saturation regime the second term in brackets can be neglected in comparison
with unity only under the condition $R_p\ll R $. Thus, we see that saturating absorption significantly affects the rate of phase diffusion.

The expression (\ref{43}) is interpreted as phase diffusion and should be treated as follows. If we performed a measuring procedure with respect to
the phase, then we will find some specific value of the lasing phase. However, due to phase diffusion, the lasing phase changes in the peculiar time
$\langle \Phi_\varphi^2\rangle^{-1} $ and becomes unpredictable.

\section{Quadrature squeezing of a laser field with a saturated absorption under the phase synchronization conditions}

Let us return to the Eq. (\ref{23}), assuming now that the laser radiation is synchronized by an external signal with amplitude
$a_{in}=\sqrt{n_{in}}\;\exp(i\varphi_{in})$. It can be expected that the fluctuations of the complex amplitude compared to the stationary amplitude
$\tilde a$ (see (\ref{28})) are small. This distinguishes the considered case from the previous one, where we followed the number of photons, and the
phase fluctuations were not limited at all. Let us set
\BY
&&a(t) =\tilde a+\delta a(t),\L{45}\\
&& \mbox{Re}\[\tilde a\]\gg |\mbox{Re}\[\delta a(t)\]|,\qquad \mbox{Im}\[\tilde a\]\gg |\mbox{Im}\[\delta a(t)\]|.\nn
\EY
Here, the stationary amplitude $\tilde a$ is given by
\BY
&&\kappa\;(1-\mu)= \frac{A}{1+\beta |\tilde a|^2}-\frac{A_p}{1+\beta_p |\tilde a|^2},\L{46}\\
&&\mu = \sqrt{n_{in}/\tilde n}.\nn
\EY
The main interest for us is not the complex amplitude itself, but its quadratures. Fluctuations of quadratures in a convenient basis can be written
in the form:
\BY
&&\delta x=\frac{1}{2}\(e^{-i\varphi_{in}}\delta a+e^{i\varphi_{in}}\delta a^\ast\),\L{47a}\\
&&\delta y=\frac{1}{2i}\(e^{-i\varphi_{in}}\delta a-e^{i\varphi_{in}}\delta a^\ast\).\L{47b}
 \EY
Due to the linearization of the equation (\ref{23}), for these quadratures we obtain two independent equations:
\BY
&& \dot{\delta x} =-(\kappa\mu/2+D)\;\delta x +\Phi_x,\L{48a}\\
&& \dot{\delta y} =-\kappa\mu/2\;\delta y +\Phi_y.\L{48b}
 \EY
Here the coefficient $ D $ is determined by the expression:
\BY
D=\kappa (1-\mu)\;\frac{\tilde I}{1+\tilde I}+\frac{A_p}{1+\tilde I_p}\(\frac{\tilde I}{1+\tilde I}-\frac{\tilde I_p}{1+\tilde I_p}\)\L{55} .
\EY
Langevin sources in the equations (\ref{48a}) - (\ref{48b}) are related to the sources $\Phi $ and $\Phi_p $ as following:
\BY
 \Phi_x=&&\frac{1}{2}\[e^{-i\varphi_{in}}\Phi+ e^{i\varphi_{in}}\Phi^\ast\]\nn\\
&& +\frac{1}{2}\[e^{-i\varphi_{in}}\Phi_p+ e^{i\varphi_{in}}\Phi^\ast_p\],\L{49}\\
 \Phi_y=&&\frac{1}{2i}\[e^{-i\varphi_{in}}\Phi- e^{i\varphi_{in}}\Phi^\ast\]\nn\\
&& +\frac{1}{2i}\[e^{-i\varphi_{in}}\Phi_p- e^{i\varphi_{in}}\Phi^\ast_p\].\L{50}
\EY
Further we will discuss the problem in terms of Fourier components, then the equations (\ref{48a}) - (\ref{48b}) will turn into equalities
\BY
&& -i\omega \delta x_\omega
 =-(\kappa\mu/2+D)\;\delta x_\omega
 +\Phi_{x \omega},\L{51a}\\
&& -i\omega\delta y_\omega
 =-\kappa\mu/2\;\delta y_\omega +\Phi_{y\omega}.\L{51b}
 \EY
Then, taking into account the correlation relations for $\Phi$ and $\Phi_p$ (\ref{24}) - (\ref{25}), we can obtain the following:
\BY
&& \langle|\Phi_{x\omega}|^2\rangle=D_1/2,\L{52a}\\
&&\langle|\Phi_{y\omega}|^2\rangle=\kappa/2(1-\mu)+ \frac{A_p}{1+\tilde I_p},\L{52b}
\EY
where the coefficient $D_1$ is given by
\BY
 D_1=&&\kappa(1-\mu)\;\frac{1-s/2\;\tilde I}{1+\tilde I}\nn\\
&&+ \frac{A_p}{1+\tilde I_p}\(\frac{1-{s}/{2}\;\tilde I}{1+\tilde I}+\frac{1}{1+\tilde I_p}\).
\EY

Knowing the spectral noise powers (\ref{52a}) - (\ref{52b}), one can explicitly write down the spectral powers of the corresponding quadrature
components of the intracavity field, and using the well-known input-output relation, connect them with those observed outside the cavity.

To observe quadratures, the balanced homodyne detection is usually applied. In this approach, a signal with an amplitude $ \hat a_{out}(t) $,
released from the cavity, before getting to the photodetector, is mixed with a classical field of local oscillator $L(t)$. As a result, the
photocurrent fluctuation operator is written as
\BY
&&\delta\hat i(t)=L^\ast(t)\delta \hat a_{out}(t)+L(t)\delta \hat
a_{out}^\dag(t).\L{53}
\EY

We can follow the quadrature of interest to us by choosing an appropriate local oscillator. For example, choosing $L=|L|\exp(i\varphi_{in})$, we will
follow the $ x $-quadrature: $\delta \hat i(t)= 2|L|\delta x_{out}(t)$. To observe the $y$-quadrature, you need to shift the phase of the local
oscillator $\varphi_{in}$ by $\pi/2$.

The $x$-quadrature of the field is of the greatest interest. With regard to the previous formulas in this section, we obtain the photocurrent noise
spectrum in the form
\BY
&&\langle|\delta i_\omega|^2\rangle/|{L}|^2=1+ \frac{2\kappa D_1}{(D+\kappa\mu/2)^2+\omega^2}\quad.\L{54}
\EY
In the regime of saturation of both media $ D =\kappa(1-\mu),\; D_1=-s/2 \;\kappa(1-\mu) $. It is not difficult to see that under these conditions
\BY
&&\langle|\delta i_{\omega=0}|^2\rangle/|L|^2=\mu^2/4\ll1.\L{55}
\EY
We can conclude  that the additional cell inserted into the laser cavity does not lead to the demolition of the quantum features of the
sub-Poissonian laser, regardless of whether it is synchronized with an external field or not.

\section{Summary}

In this paper, we have shown that an additional cell carrying out a saturated absorption does not lead to the demolition of the quantum features of a
sub-Poissonian laser, regardless of whether it is synchronized with an external field or not. This applies to both quantum effects, namely, photon
statistics and field quadratures squeezing.

A regular pump of a medium in a laser leads to the fact that the statistics of photons in the saturation regime turns out to be sub-Poissonian. With
a gradual decrease in gain, quantum features are becoming less and less expressive. This is due to the fact that at low generation powers, part of
the radiation is scattered isotropically on the medium in the form of spontaneous radiation. This forms uncontrollable losses, which lead to the
destruction of the quantum features of radiation.

The appearance in the resonator of an additional element in the form of an absorbing cell also leads to uncontrolled losses associated with
spontaneous scattering by the atoms of the passive medium. When the absorbing medium is saturated, uncontrolled losses disappear. Thus, in the
limiting case of $\tilde I, \; \tilde I_p\gg1 $, the quantum properties of a sub-Poissonian laser are not subject to damage by an additional cell
that performs saturated absorption.

At the same time, it may also be of interest that the phase diffusion rate in a laser with an absorbing cell turns out to be significantly increased
compared to a laser without a cell.

\section{Acknowledgements}

This work was supported by RFBR (grants 18-02-00402a, 18-02-00648a and 16-02-00180a) and by grant on Nonlinear dynamics of Presidium of the Russian
Academy of Sciences. Scientific results were achieved during the implementation of the Program within the state support of the STI Center "Quantum
Technologies".

\end{document}